\documentstyle[12pt]{article}
\begin{document}
\large
\begin{centerline}
{\bf PROBLEMS WITH NONPERTURBATIVE EFFECTS}
\end{centerline}
\begin{centerline}
{\bf IN DEEP INELASTIC SCATTERING}
\end{centerline}

 \begin{center} Felix M. Lev \end{center}

\begin{center}
{\it Laboratory of Nuclear Problems, JINR, Dubna, Moscow region 141980,
 Russia (E-mail: lev@nusun.jinr.dubna.su)}
\end{center}

\begin{abstract}
We consider restrictions imposed on the (electromagnetic or weak)
current operator by its commutation relations with the
representation operators of the Poincare group  and show that
the nonperturbative part of the current operator contributes to
deep inelastic scattering even in leading order in $1/Q$ where $Q$ is
the magnitude of the momentum transfer. Some consequences of this
result are discussed.

\begin{flushleft} PACS: 11.40.Dw, 13.60.Hb \end{flushleft}

\begin{flushleft} Key words: deep inelastic scattering, nonperturbative
effects, operator product expansion. \end{flushleft}

\end{abstract}

 {\bf 1.} The present theory of deep inelastic scattering (DIS) is based
on two approaches which are the complement of one another. In the first
approach (see e.g. ref. \cite{LP} and references therein) one assumes
that only Feynman diagrams from a certain class dominate in DIS, and in
the second approach DIS is considered in the
framework of the operator product expansion (OPE) \cite{Wil}.
Although the assumptions used in the both approaches are natural, the
problem of their substantiation remains since we do not know how to
work with QCD beyond perturbation theory. In particular, the OPE has
been proved only in perturbation theory \cite{Br} and its validity beyond
that theory is problematic (see the discussion in ref. \cite{Nov} and
references therein).

  In the present paper we show that an important information about the
structure of the (electromagnetic or weak) current operator in DIS can
be obtained from the investigation of restrictions imposed on this
operator by its commutation relations with the representation operators
of the Poincare group.

  {\bf 2.} If $J(x)$ is the electromagnetic or weak current
operator and $q$ is the momentum transfer then the DIS cross-section
is fully defined by the hadronic tensor
\begin{equation}
W^{\mu\nu}=\frac{1}{4\pi}\int\nolimits e^{\imath qx} \langle P'|
J^{\mu}(x)J^{\nu}(0)|P'\rangle d^4x
\label{1}
\end{equation}
where $|P'\rangle$ is the state of the initial nucleon with the
four-momentum $P'$ and we use $\mu,\nu=0,1,2,3$ to denote the
components of the operator $J(x)$.

 Translational invariance of the current operator implies that
 \begin{equation}
J(x)=exp(\imath Px) J(0)exp(-\imath Px),
\label{2}
\end{equation}
where $P$ is the four-momentum operator, and Lorentz invariance
implies that
\begin{equation}
[M^{\mu\nu},J^{\rho}(0)]=-\imath (g^{\mu\rho}J^{\nu}(0)-g^{\nu\rho}
J^{\mu}(0))
\label{3}
\end{equation}
where $M^{\mu\nu}$ are the Lorentz group generators and $g^{\mu\nu}$
is the Minkowski tensor.

 In turn, the state $|P'\rangle$ is the eigenstate of the operator $P$
with the
eigenvalue $P'$ and the eigenstate of the spin operators ${\bf S}^2$ and
$S^z$ which are constructed from $M^{\mu\nu}$. In particular,
$P^2|P'\rangle =m^2 |P'\rangle$ where $m$ is the nucleon mass. Therefore
the four-momentum operator necessarily depends on the soft part of the
interaction which is responsible for binding of quarks and gluons in
the nucleon. The Lorentz transformations of the nucleon state are described
by the operators $M^{\mu\nu}$ and therefore in the general case they
also depend on the soft part.

 It is important to note that the same operators $(P^{\mu},M^{\mu\nu})$
describe the transformations of both the operator $J(x)$ and the
state $|P'\rangle$, and this guaranties that $W^{\mu\nu}$ has the
correct transformation properties.

 We see that the relation between the current operator and the state of
the initial nucleon is highly nontrivial. Meanwhile in the present theory
they are considered separately. In the framework of the approach based
on Feynman diagrams the possibility of the separate consideration
follows from the factorization theorem \cite{ER} which asserts in
particular that the amplitude of the lepton-parton interaction
entering into diagrams dominating in DIS depend only on the hard part
of this interaction. Moreover, in leading order in $1/Q$,
where $Q=|q^2|^{1/2}$, one obtains the parton model up
to anomalous dimensions and perturbative QCD corrections which depend on
$\alpha_s(Q^2)$ where $\alpha_s$ is the QCD running coupling constant.

 It is well-known that the parton model is equivalent to impulse
approximation (IA) in the infinite momentum frame (IMF). This fact is in
agreement with our experience in conventional nuclear and atomic physics
according to which in processes with high momentum transfer the effect of
binding is not important and the current operator can be taken in IA.
However this experience is based on the nonrelativistic
quantum mechanics where only the Hamiltonian is interaction dependent
and the other nine generators of the Galilei group are free. Note
also that in the nonrelativistic case the kinetic energies and the
interaction operators in question are much smaller than the masses of
the constituents.

 The usual motivation of the parton model is that,
as a consequence of asymptotic freedom (i.e. the fact
that $\alpha_s(Q^2)\rightarrow 0$ when $Q^2 \rightarrow \infty$),
the partons in the IMF are almost free and therefore, at least in
leading order in $1/Q$, the soft part of $J(x)$ is not important. We will
consider a bit later whether this property can be substantiated in the
framework of the OPE but first we consider some consequences of Eqs.
(2) and (3).

 {\bf 3.}  As noted by Dirac \cite{Dir}, the operators
$(P^{\mu},M^{\mu\nu})$ can
be realized in different representations, or, in Dirac's terminology,
in different forms of dynamics. Suppose that the Hamiltonian $P^0$
contains the soft part and consider the well-known relation
$[M^{0i},P^k]=-\imath \delta_{ik}P^0$ $(i,k=1,2,3)$. Then it is obvious
that if all the operators $P^k$ are free then all the operators $M^{0i}$
inevitably contain the soft part and {\it vice versa},
if all the operators $M^{0i}$ are free then all the operators $P^k$
inevitably contain this part. According to the Dirac classification
\cite{Dir}, in the instant form the Hamiltonian
$P^0$ and the operators $M^{0i}$  are interaction dependent
and the other six generators of the Poincare group are free, while
in the point form all the components $P^{\mu}$
are interaction dependent and all the operators $M^{\mu\nu}$ are free.
In the front form the operators $P^-$ and $M^{-j}$ ($j=1,2$,
$p^{\pm}=p^0\pm p^z$) are interaction dependent and the other seven
generators are free. The fact that if $P^-$ is the only dynamical
component of $P$ then all the $M^{-j}$
inevitably contain interaction terms follows from the relation
$[M^{-j},P^l]=-\imath \delta_{jl}P^-$. Of course, the physical results
should not depend on the choice of the form of dynamics and in the
general case all ten generators can be interaction dependent.

\begin{sloppypar}
  The usual form of the electromagnetic current operator is
$J^{\mu}(x)={\cal N}\{{\bar \psi}(x)\gamma^{\mu}\psi (x)\}$ and in
particular $J^{\mu}(0)={\cal N}\{{\bar \psi}(0)\gamma^{\mu}\psi (0)\}$,
where ${\cal N}$ stands for the normal product and for simplicity we do
not write flavor operators and color and flavor indices.
However such a definition ignores the fact that the product of two
field operators at coinciding points is not a well-defined operator
(strictly speaking, the operator $\psi (0)$ also is not defined
since $\psi (x)$ is the operator-valued distribution; for a more
detailed discussion see ref. \cite{hep}). The reader
thinking that it is not reasonable to worry about the mathematical
rigor will be confronted with the following contradiction.
\end{sloppypar}

 The canonical quantization on the hyperplane $x^0=0$ or on the light
cone $x^+=0$ (which leads to the instant and front forms respectively
\cite{Dir}) implies that the operator $\psi (0)$ is free since the
Heisenberg and Schrodinger pictures coincide at $x=0$. Then $J(0)$
is free too and, as follows from Eq. (3), the interaction terms in
$M^{\mu\nu}$ should commute with $J^{\rho}(0)$. If the operators
$M^{\mu\nu}$ are constructed by means of canonical quantization then in
QED the interaction terms and their commutators with $J^{\rho}(0)$
can be readily calculated. The commutators are expressed in
terms of the Schwinger terms \cite{Schw} which cannot be equal to zero
(the corresponding calculation is given in ref. \cite{hep}).
Therefore the conclusion that all the components of $J(0)$ are free is
incorrect and some components of $J(0)$ are inevitably interaction
dependent.

\begin{sloppypar}
 Moreover, it can be shown that if the field operators are
quantized, for example, on the hyperplane $x^0=0$ then the operator
${\bf J}(0)$ in QED is necessarily interaction dependent.
Indeed, the generator of the gauge transformations is
$div {\bf E}({\bf x}) - J^0({\bf x})$, and if ${\bf J}(0)$ is gauge
invariant then $[div {\bf E}({\bf x}) - J^0({\bf x}),
{\bf J}(0)]=0$. The commutator $[J^0({\bf x}),{\bf J}(0)]$ cannot be
equal to zero \cite{Schw} and therefore $J^0({\bf x})$ does not
commute with $div {\bf E}({\bf x})$ while the free operator
$J^0({\bf x})$ commutes with $div {\bf E}({\bf x})$.
\end{sloppypar}

 The above examples illustrate the well-known fact that formal
manipulations with local operators in quantum field theory can lead to
incorrect results. For this reason we prefer to rely only upon algebraic
considerations according to which all the components of $J(0)$
cannot be free simply because there is no reason for the interaction
terms in $M^{\mu\nu}$ to commute with the free operators $J^{\rho}(0)$
(see Eq. (3)). Therefore in the instant and front forms
some of the operators $J^{\rho}(0)$ depend on the soft part.
On the other hand, as follows from Eq. (3), if the operator $J(0)$
is free in the point form, this does not contradict Lorentz invariance
but, as follows from Eq. (2), the operator $J(x)$ in that form
necessarily contains the soft part.

 The problem of the correct definition of the product of two local
operators at coinciding points is known as
the problem of constructing the composite operators (see e.g. ref.
\cite{Zim}). So far this problem has been solved only in the framework
of perturbation theory for special models. When perturbation theory
does not apply the usual prescriptions are to separate the arguments
of the operators in question and to define the composite operator as
a limit of nonlocal operators when the separation goes to zero (see e.g.
ref. \cite{J} and references therein). Since we do not know how to
work with quantum field theory beyond perturbation theory, we do not
know what is the correct prescription. Moreover, it is not clear at all
whether it is possible to define local interaction dependent operators
in QCD. Indeed, the dependence of an operator on the soft part implies
that the operator depends on the integrals from the quark and gluon
field operators over the region of large distances where the QCD
running coupling constant $\alpha_s$ is large. It is obvious that such
an operator cannot be local. In particular it is not clear whether
in QCD it is possible to construct local electromagnetic and weak current
operators beyond perturbation theory.

 {\bf 4.} In the framework of the OPE the product of the currents
entering into Eq. (1) can be written symbolically as
\begin{equation}
J(x)J(0) = \sum_{i} C_i(x^2) x_{\mu_1}\cdots x_{\mu_n}
O_i^{\mu_1\cdots \mu_n}
\label{4}
\end{equation}
where $C_i(x^2)$ are the $c$-number Wilson coefficients while the
operators $O_i^{\mu_1\cdots \mu_n}$ depend only on field operators and
their covariant derivatives at the origin of Minkowski space and have
the same form as in perturbation theory. The basis for twist two operators
contains in particular
\begin{equation}
O_V^{\mu}={\cal N}\{{\bar \psi}(0)\gamma^{\mu}\psi (0)\}, \quad
O_A^{\mu}={\cal N}\{{\bar \psi}(0)\gamma^5\gamma^{\mu}\psi (0)\}
\label{5}
\end{equation}

 As noted above, the operator $J(x)$ necessarily depends on the
soft part while Eq. (4) has been proved only in the framework of
perturbation theory. Therefore if we use Eq. (4) in DIS we have to
assume that either nonperturbative effects are not important to some
orders in $1/Q$ and then we can use Eqs. (1) and (4) only to these
orders (see e.g. ref. \cite{Jaffe}) or it is possible to use Eq. (4)
beyond perturbation theory. The question also arises whether Eq. (4)
is valid in all the forms of dynamics (as it should be if it is an
exact operator equality) or only in some forms.

 In the point form all the components of $P$ depend on the soft part
and therefore, in view of Eq. (2), it is not clear why there is no
soft part in the $x$ dependence of the right hand side of Eq. (4),
or if it is possible to include the soft part only into the operators
$O_i$ then why they have the same form as in perturbation theory.

 One might think that in the front form the $C_i(x^2)$ will be the
same as in perturbation theory due to the following reasons. The
value of $q^-$ in DIS is very large and therefore only a small
vicinity of the light cone $x^+=0$ contributes to the integral (1).
The only dynamical component of $P$ is $P^-$ which enters into
Eq. (4) only in the combination $P^-x^+$. Therefore the dependence
of $P^-$ on the soft part is of no importance.  These considerations
are not convincing since the integrand is a singular function and the
operator $J(0)$ depends on the soft part in the front form, but
nevertheless we assume that Eq. (4) in the front form is valid.

 If we assume as usual that there is no problem with the convergence
of the OPE series then experiment makes it possible to measure
each matrix element $\langle P'|O_i^{\mu_1\cdots \mu_n}|P'\rangle$.
Let us consider, for example, the matrix element
$\langle P'|O_V^{\mu}|P'\rangle$. It transforms as a four-vector if the
Lorentz transformations of $O_V^{\mu}$ are described by the operators
$M^{\mu\nu}$ describing the transformations of $|P'\rangle$, or in
other words, by analogy with Eq. (3)
\begin{equation}
[M^{\mu\nu},O_V^{\rho}]=-\imath (g^{\mu\rho}O_V^{\nu}-g^{\nu\rho}
O_V^{\mu})
\label{6}
\end{equation}
It is also clear that Eq. (6) follows from Eqs. (2-4).
Since the $M^{-j}$ in the front form depend on the soft part,
we can conclude by analogy with the above consideration that at least
some components $O_V^{\mu}$, and analogously some components
$O_i^{\mu_1\cdot \mu_n}$, also depend on the soft part. Since Eq. (6)
does not depend on $Q$, this conclusion has nothing to do with asymptotic
freedom and is valid even in leading order in $1/Q$ (in contrast with the
statement of the factorization theorem \cite{ER}).
Since the struck quark is not free but interacts nonperturbatively with
the rest of the target then, in terminology of ref. \cite{LP}, not
only "handbag" diagrams dominate in DIS but some "cat ears" diagrams
or their sums are also important (in other words, even the notion of
struck quark is questionable).

 Since the operators $O_i^{\mu_1...\mu_n}$ depend on the soft part
then by analogy with the considerations in subsection 3 we conclude
that the operators in Eq. (5) are ill-defined and the
correct expressions for them involve integrals from the
field operators over large distances where the QCD coupling constant is
large. Therefore the Taylor expansion at $x=0$
is questionable, and, even if it is valid, the expressions for
$O_i^{\mu_1...\mu_n}$ will depend on higher twist operators which
contribute even in leading order in $1/Q$.

 {\bf 5.} Let us now discuss our results. First we have shown that
the current operator nontrivially depends on the nonperturbative part
of the interaction responsible for binding of quarks and gluons in the
nucleon. Then the problem arises whether this part contributes to DIS.
Our consideration shows that the dependence of $J(x)$ on the
nonperturbative part of the interaction makes the OPE problematic.
Nevertheless we assume that Eq. (4) is valid beyond perturbation
theory but no form of the operators $O_i^{\mu_1...\mu_n}$ is prescribed.
Then we come to conclusion that the nonperturbative part contributes to
DIS even in leading order in $1/Q$.

 To understand whether the OPE is valid beyond perturbation theory
several authors (see e.g. ref. \cite{Nov} and references therein)
investigated some two-dimensional models and came to different
conclusions. We will not discuss the arguments of these authors but
note that the Lie algebra of the Poincare group for 1+1 space-time
is much simpler than for 3+1 one. In particular, the Lorentz group is
one-dimensional and in the front form the operator $M^{+-}$ is free.
Therefore Eqs. (3) and (6) in the "1+1 front form" do not make it
possible to conclude that the operators $J^{\rho}(0)$ and $O_V^{\rho}$
should depend on the nonperturbative part of the quark-gluon interaction.

 Since the operators $O_i^{\mu_1...\mu_n}$ in Eq. (4) should depend
on the nonperturbative part of the quark-gluon interaction then,
as noted above, there is no reason to think that these operators
are local but even if they are then twist (dimension minus spin) no
longer determines in which order in $1/Q$ the corresponding
operator contributes to DIS. This is clear from the fact that the
dependence on the nonperturbative part implies that we have an
additional parameter $\Lambda$ with the dimension of momentum
where $\Lambda$ is the characteristic momentum at which
$\alpha_s(\Lambda^2)$ is large.

 Nevertheless if we assume that (for some reasons) Eq. (4) is still
valid and consider only the $q^2$ evolution of the structure functions
then all the standard results remain. Indeed the only information
about the operators $O_i^{\mu_1...\mu_n}$ we need is their tensor
structure since we should correctly parametrize the matrix elements
$\langle P'|O_i^{\mu_1\cdots \mu_n}|P'\rangle$. However the
derivation of sum rules in DIS requires additional assumptions.

 Let us consider sum rules in DIS in more details. It is well-known
that they are derived with different extent of rigor. For example,
the Gottfried and Ellis-Jaffe sum rules \cite{Got} are essentially
based on model assumptions, the sum rule \cite{Adl} was originally
derived in the framework of current algebra for the time component
of the current operator while the sum rules \cite{Bjor} also involve
the space components. As noted in subsection 3, the operator
${\bf J}(0)$ is necessarily interaction dependent; on the other hand
there exist models in which $J^0(0)$ is free (see e.g. calculations
in scalar QED in ref. \cite{hep}). Therefore in the
framework of current algebra the sum rule \cite{Adl} is substantiated
in greater extent than the sum rules \cite{Bjor} (for a detailed
discussion see refs. \cite{GM,J}). Now the sum rules \cite{Adl,Bjor}
are usually considered in the framework of the OPE and they
have the status of fundamental relations which in fact unambiguously
follow from QCD. However the important assumption in deriving the sum
rules is that the expression for $O_V^{\mu}$ coincides with
$J^{\mu}(0)$, the expression for $O_A^{\mu}$ coincides with the axial
current operator $J_A^{\mu}(0)$ etc. (see Eq. (5)). Our results
show that this assumption has no physical ground. Therefore although
(for some reasons) there may exist sum rules which are satisfied with
a good accuracy, the statement that the sum rules \cite{Adl,Bjor}
unambiguously follow from QCD is not substantiated.

 For comparing the theoretical predictions for the sum rules with
experimental data it is also very important to calculate effects in
next-to-leading order in $1/Q$. As shown in ref. \cite{Mart} there
exist serious difficulties in calculating such effects in the
framework of the OPE, and the authors of ref. \cite{Mart} are very
pessimistic about the possibility to overcome these difficulties
(while in our approach problems exist even in the leading order).

 The current operator satisfying Eqs. (2) and (3) can be explicitly
constructed for systems with a fixed number of interacting
relativistic particles \cite{lev}. In such models it is clear when
the corresponding results and the results in IA are similar and when
they considerably differ \cite{hep1}.

 We conclude that the present theory of DIS based on perturbative
QCD does not take into account the dependence of the current operator
on the nonperturbative part of the quark-gluon interaction which
cannot be neglected even in leading order in $1/Q$. On the other
hand, the present theory has proven rather successful in describing
many experimental data. It is very important to understand why this
situation takes place.

\begin{sloppypar}
 {\it Acknowledgments.} The author is grateful to G.Salme and E.Pace
for the idea to write this paper, constructive criticism
and valuable discussions. The author has also greatly benefited from
the discussions with B.L.G.Bakker, R. Van Dantzig, A.V.Efremov,
L.A.Kondratyuk, B.Z.Kopeliovich, M.P.Locher, M.Marinov, S.Mikhailov,
P.Muelders, N.N.Nikolaev, R.Petronzio, R.Rosenfelder,
O.Yu.Shevchenko, I.L.Solovtsov and H.J.Weber. This work is supported
by grant No. 96-02-16126a from the Russian Foundation for Fundamental
Research.
\end{sloppypar}


\begin{thebibliography}{99}
\bibitem{LP} P.V.Landshoff and J.C.Polkinghorne, Phys. Rep. 5 (1972) 1;
F.Close, {\it An Introduction to Quark and Partons} (Academic Press,
London - New York - San Francisco, 1979).
\bibitem{Wil} K.G.Wilson, Phys.Rev. 179 (1969) 1499; Phys. Rev.
D 3 (1971) 1818.
\bibitem{Br} R.Brandt, Fortschr. Phys. 18 (1970) 249;
W.Zimmerman, Annals of Physics 77 (1973) 570.
\bibitem{Nov} V.A.Novikov, M.A.Shifman, A.I.Vainshtein and V.Zakharov,
Nucl. Phys. B 249 (1985) 445;
M.Soldate, Annals of Physics 158 (1984) 433;
F.David, Nucl. Phys. B 263 (1986) 637.
\bibitem{ER} A.V.Efremov and A.V.Radyushkin, Rivista Nuovo Cimento
3, No 2, (1980); G.Curci, W.Furmanski and R.Petronzio, Nucl. Phys.
B 175 (1982) 27; R.K.Ellis, W.Furmanski and R.Petronzio, Nucl. Phys.
B 212 (1983) 29.
\bibitem{Dir} P.A.M.Dirac, Rev. Mod. Phys. 21 (1949) 392.
\bibitem{hep} F.M.Lev, hep-ph 9512010.
\bibitem{Schw} J.Schwinger, Phys. Rev. Lett. 3 (1959) 296.
\bibitem{Zim} W.Zimmerman, Annals of Physics 77 (1973) 536;
O.I.Zavialov. {\it Renormalized Feynman Diagrams} (Nauka,
Moscow 1979).
\bibitem{J} R.Jackiw, in {\it Lectures on Current Algebra and its
Applications}, S.Treiman, R.Jackiw and D.Gross, eds. (Princeton
University Press, Princeton NJ 1972).
\bibitem{Jaffe} R.L.Jaffe and M.Soldate, Phys. Lett. B 105 (1981) 467.
\bibitem{Got} K.Gottfried, Phys. Rev. Lett. 18 (1967) 1174;
J.Ellis and R.L.Jaffe, Phys. Rev. D 9 (1974) 1444, D 10
(1974) 1669.
\bibitem{Adl} S.L.Adler, Phys. Rev. 143 (1966) 1144.
\bibitem{Bjor} J.D.Bjorken, Phys. Rev. 148 (1966) 1467, 163
(1967) 1767; D.J.Gross and C.H.Llewellyn Smith, Nucl. Phys. B 14 (1969)
337.
\bibitem{GM} M.Gell-Mann, Physics 1 (1964) 63.
\bibitem{Mart} G.Martinelli and C.T.Sachrajda, hep-ph 9605336.
\bibitem{lev} F.M.Lev, Annals of Physics 237 (1995) 355.
\bibitem{hep1} F.M.Lev, hep-ph 9507421, 9511406.
\end{thebibliography}
\end{document}